# DistB-Condo: Distributed Blockchain-based IoT-SDN Model for Smart Condominium

ANICHUR RAHMAN[1,2], MD. JAHIDUL ISLAM[3], ZIAUR RAHMAN[4], MD. MAHFUZ REZA[1], ADNAN ANWAR[5], M. A. PARVEZ MAHMUD[6], MOSTOFA KAMAL NASIR[1], and RAFIDAH MD NOOR[7,8]

[1]Department of Computer Science and Engineering, Mawlana Bhashani Science and Technology University, Tangail, Bangladesh
[2]Department of Computer Science and Engineering, National Institute of Textile Engineering and Research (NITER), Savar, Dhaka, Bangladesh
[3]Green University of Bangladesh, Dhaka, Bangladesh
[4]Department of Information and Communication Technology, Mawlana Bhashani Science and Technology University, Tangail, Bangladesh
[5]Deakin University, Melbourne VIC 3220, Australia
[6]School of Engineering, Deakin University, Geelong, VIC 3216, Australia
[7]Faculty of Computer Science and Information Technology, University of Malaya, 50603 Kuala Lumpur, Malaysia
[8]Center for Mobile Cloud Computing Research, Faculty of Computer Science and Information Technology, University of Malaya, 50603 Kuala Lumpur, Malaysia

Corresponding author: Mostofa Kamal Nasir and Anichur Rahman

**ABSTRACT** Condominium network refers to intra-organization networks, where smart buildings or apartments are connected and share resources over the network. Secured communication platform or channel has been highlighted as a key requirement for a reliable condominium which can be ensured by the utilization of the advanced techniques and platforms like Software-Defined Network (SDN), Network Function Virtualization (NFV) and Blockchain (BC). These technologies provide a robust, and secured platform to meet all kinds of challenges, such as safety, confidentiality, flexibility, efficiency, and availability. This work suggests a distributed, scalable IoT-SDN with Blockchain-based NFV framework for a smart condominium (DistB-Condo) that can act as an efficient secured platform for a small community. Moreover, the Blockchain-based IoT-SDN with NFV framework provides the combined benefits of leading technologies. It also presents an optimized Cluster Head Selection (CHS) algorithm for selecting a Cluster Head (CH) among the clusters that efficiently saves energy. Besides, a decentralized and secured Blockchain approach has been introduced that allows more prominent security and privacy to the desired condominium network. Our proposed approach has also the ability to detect attacks in an IoT environment. Eventually, this article evaluates the performance of the proposed architecture using different parameters (e.g., throughput, packet arrival rate, and response time). The proposed approach outperforms the existing OF-Based SDN. DistB-Condo has better throughput on average, and the bandwidth (Mbps) much higher than the OF-Based SDN approach in the presence of attacks. Also, the proposed model has an average response time of 5% less than the core model.

## I. INTRODUCTION

LATELY, in the concepts of condominium, people live together with their apartments where some shared area satisfies their demand. Condominium causes re-urbanization and re-evolution of the inner city [1]. A smart community may take the advantage of controlling the resources within a condominium to improved the quality of lifestyle, and specifically, it is flexible while the control and decision making system is designed based on cloud architecture [2]. In a typical setup, the resources are shared over a network as all the people of a condominium can't have all the resources on their own such as car parking areas, water resources, security systems, and so on. Moreover, to lead a felicitous life, there should be proper distribution of the shared resources among the people living in a condominium. The IoT is used for digitizing the housing system like managing and controlling the resources (light, fans, even security systems) digitally through computing and communication. The processes that deal with digital housing can use the approaches of IoT [3]. In daily life, IoT is used for its various beneficial aspects [4]. IoT devices can send information over the network, which is the primary criterion of a smart condominium. Different IoT architectures have been used for smart cities, and the architectures premise the protocols used in IoT network [5]. Besides, SDN offers a various gateway to analyze traffic patterns through data transmission over the IoT network [6]. Therefore, integration of IoT and SDN makes the architecture flexible and programmable.

Although the above discussed IoT and SDN based architecture has opened the door of enormous potential, it is centralized in nature and uses OpenFlow protocols to perform different operations. On the contrary, Blockchain is another data structure that provides transactional financial security into the system through a digital ledger of transactions, and it shares the ledger by utilizing a distributed network. The process of exchange in the shared resources can use Blockchain in the area of smart condominium and collaborative mechanisms [7]. Further, Network Function Virtualization (NFV) virtualizes the SDN controller and replaces the network appliances like switches, routers, firewalls, etc. with software [8]. The architecture of NFV is made with virtual machines performing different processes. Blockchain, SDN, NFV could be used in a combination to provide a fabulous abstraction and performance into the establishment of smart condominium [9].

A smart Condo (condo is the shortened form of condominium that we will be using henceforth) was deployed by the researchers of the University of Alberta [10] through a wireless sensor network. They emphasize on the afford- able healthcare and visualizing the collected information. The sensors monitored the activities, and the condominium system received the information. Moreover, re-urbanization was performed using the condominium boom and social housing. As the buildings and multi-unit buildings in a city are increasing with time, it became a challenge to ensure their security and management. The infrastructure of a smart condominium needs to be more specific and efficient, and the whole management should be energy saving. However, it also poses some challenges, specifically related to the IoT security as highlighted by Bouanani et al. [11]. The security issues of IoT is still an open research area and ongoing efforts are made by the researchers, scientists, technologists to mitigate these challenges related to security and data sharing. Blockchain has the potential to access the housing information [12] towards effective data transactions.

For secure data transmission, it is challenging to create a protocol that ensures efficiency, QoS, and Privacy. Moreover, the above technologies are still under development and getting matured over the time with recent progresses [13]. This research proposes an architecture for a smart condo utilizing distributed Blockchain based IoT-SDN. It enables distributed data transfer, sharing and decision making. Moreover, the most significant part of this architecture is to provide security among highly interconnect and interrelated smart communities. For these reasons, the authors have used blockchain and SDN technologies that ensure secured transmission of data and effective resource management. Besides, smart condominium based on SDN includes many significant components such as system scope, target architecture, network controller, communication protocols, creativity setup, etc. The aim of this work is to ensure protection, safety, zip performance, the path of the network, reliability, QoS and reduced latency within a smart condo.

The paper contributions are as follows:

- Design & development of a distributed Blockchain-SDN based secure "DistB-Condo" architecture, which provides security and privacy within a building and more essential data confidentiality from one community to another in the smart condominium applications.
- Present a CHS algorithm that picking head rapidly with the lower dissipation of energy from IoT devices.
- Also uses SDN controllers to monitor and manage IoT data traffics; Blockchain is used to detect & mitigate the security attacks and secure data transaction.
- Finally, the evaluation of result indicates that our proposed model provides significantly improved efficiency of security compared to the existing works with minimized response time and higher throughput using the blockchain technology.

Based on the contributions, we highlight the Blockchain network regarding secured transaction purposes through a smart contract environment. Then, the article focuses on SDN implementation through IoT networks. After that, we analyze the performance of the IoT-SDN network environment in different parameters like throughput, packet arrival time, file operation, etc. It can be performed using a distributed ledger with a smart contact approach to enhance the overall performances of the network efficiently.

This article is organized as follows: Authors discuss the related works and background knowledge in section II. Next, the proposed "DistB-Condo" system for an intelligent condominium application is presented in section III. The results from the experiment and a detailed discussion are shown in the following section IV. Furthermore, the future scope has been presented in section V. Besides, the authors conclude the article in the last section, which is section VI, including the significance, limitations, and future research directions.

## II. BACKGROUND AND RELATED WORKS

Recently, some researchers have reviewed the prospects of the leading technologies like SDN, Blockchain, IoT, and NFV, and proposed some technical works based on these technologies. The following section summarizes the moti-vation towards the adoption of IoT, SDN and Blockchan technologies, followed by a section with recent literature review related to these technologies. Some of the notations are listed in Table 1.

**TABLE 1.** Technical Terminologies and Description in Alphabetically Ordered

| Notations | Definition |
|---|---|
| API | Application Programming Interface |
| AWS | Amazon Web Services |
| BC | Blockchain |
| CH | Cluster Head |
| CHS | Cluster Head Selection |
| DDoS | Distributed Denial of Service |
| DHT | Distributed Hash Table |
| DPoS | Distributed Proof of Stack |
| IoT | Internet of Things |
| LEACH | Low-Energy Adaptive Clustering Hierarchy |
| LiFi | Light Fidelity |
| NFV | Network Function Virtualization |
| NFVI | Network Function Virtualization Infrastructure |
| ONOS | Open Network Operating System |
| PoW | Proof of Work |
| QoS | Quality of Service |
| RPC | Remote Procedure Call |
| SC | Smart Contact |
| SDN | Software Defined Networking |
| SDK | Software Development Kit |
| TLS | Transport Layer Security |

### A. MOTIVATION FOR IOT, SDN AND BLOCKCHAIN BASED FRAMEWORK

The main aim of the IoT is to connect smart devices for improved sensing and control. [27]. IoT includes different types of elements like peripheral devices, an Embedded system, protocols, connectivity, etc. Embedded system contains Wi-Fi, WiMax, LiFi, Bluetooth, ZigBee, Ethernet, etc. for IoT connectivity. Currently, IoT has been used in many smart technological fields such as smart healthcare, education, smart cities, as well as intelligent condominiums. In a smart condo environment, IoT offers seamless data connectivity. Moreover, a smart condo is able to communicate with each smart building through IoT based sensory data acquisition.

Although IoT interconnects multiple smart devices, there is a massive number of risks and challenges from the various types of networking model. The SDN system addresses a solution for these challenges. SDN can be able to separate into two distinct layers like infrastructure or data layer and control layers for effective management of the resources. Furthermore, it is a process of enhancing technology, which is used to evaluate and set up the network model. Then, the central aspect of SDN is to improve the performances of the network applications. Also, SDN can be capable of providing some controllers like OpenFlow, OpenDayLight, and OpenStack to manage any type of application data [20]. It can also offer some services such as privateness, accessibility, security, confidentiality, arrange the network efficiently [28]. As there is a lot of devices connected to a condominium environment, it is often challenging to ensure network connectivities and securities where SDN can play a vital role towards the smart condominium applications.

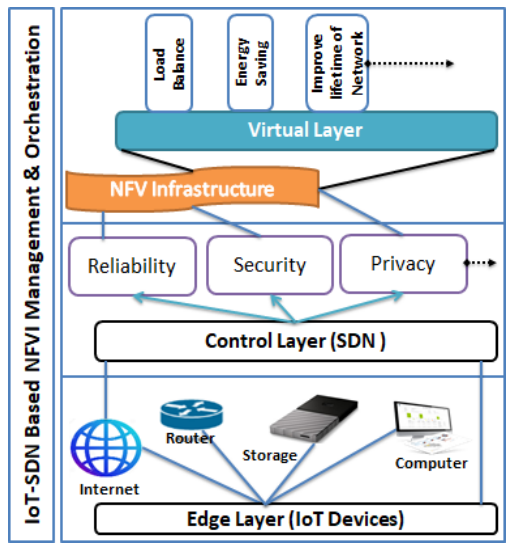

**FIGURE 1.** A notional IoT-SDN based NFVI architecture [31]

Blockchain is one of the growing technologies which works as a distributed ledger over the transactions. Basically, it was invented by Satoshi Nakamoto in 2008 to remove the double-spending problem from Bitcoin-based transaction [29]. This study uses the distributed ledger technology for the smart community, where Blockchain acts as a distributed system to enable the proposed "DistB-Condo" architecture along with SDN. Finally, this article presents IoT-SDN based architecture for NFV infrastructure, as shown in Fig. 1. This architecture is divided into several phases. In the first phase, IoT forwarding devices (router, switch, storage, etc.) can be capable of creating a dynamic group to the edge layer, and another phase (SDN platform) provides high security, high reliability, and privacy for IoT sensor data. All these data are virtualized by the virtualization technology NFV. Furthermore, the virtual layer of NFV can provide some great services named load balancing, power saving, cost-effective, and improve the lifetime to the desired smart condominium environment.

### B. RELATED WORKS

The following section summarizes some recent literature related to the IoT, SDN and Blockchain technologies for a smart environment.

#### 1) IoT Networks for Smart Applications

Plageras et al. surveyed the services of Big Data, Cloud Computing, Collection-Processing, and Internet of Things in our daily life to check out their common operations. Authors proposed an energy-efficient sensor management system combining the beneficial part of the mentioned tech- nology in the environment of IoT [4]. A cloud server and a remote device used to control the data. The approach proposed by the author was comparable to the related work. In that work, Cooja was used to simulate the proposed network. In a similar research, Sinh et al. have been concentrated on inquiring about the roles of SDN/NFV in implementing IoT services and also proposed an SDN/NFV architecture for developing IoT framework [15]. Chakrabarty et al. presented an IoT architecture for a smart city that is secure and can handle mission-critical smart city data transmitted over the IoT network [5]. Authors gave a set of components to secure the data over the large scale deployment of IoT. An extension over the basic security of IoT protocols is introduced through the proposed architecture. Authors provided privacy, identity management, and authentication with a security management system.

### 2) SDN based IoT

In [16], Matheu et al. effectively achieved the Manufacturer's Use Description (MUD) Model for Network Access Control, Data Privacy, and Channel and Authorization Protection Policies. Authors then provided SDN platforms for efficient access to device data and resources and also used Blockchain technology to share data through Hyperledger with the help of IoT devices. On the other hand, Conti et al. suggested an excellent combination of cloud, IoT, and SDN (referred as CENSOR), which could provide a secure platform for the IoT architecture [17]. They presented a stable IoT cloud-enabled model using SDN technology. The authors also presented an SDN paradigm equipped to maintain a large number of IoT data at different levels of efficiency. In another study, Kalkan et al. considered the security of various SDN inventions. Based on security demands, authors suggested the most appropriate security mechanism in [18]. The authors also discoursed future challenges in the area of IoT environment with a role-based comptroller for preserving security appointed as Rol-Sec for SDN.

### 3) IoT-SDN with NFV

Barakabitze et al. in [19] offered a review based on the SDN and NFV technology for slicing 5G networks. For a ensure Black SDN-IoT based security Islam et al. [20] aimed a distributed architecture for smart cities with the NFV implementation concept. To incorporate accessibility, security, and privacy, they also expended multiple distributed SDN controllers in their proposed architecture. In another work, Abdulqadder et al., based on SDN technology [21], presented a comprehensive security strategy over the NFV-based cloud environment in the 5g network. They have appropriately considered different risks, such as DDoS attacks, to the addressed program—the performance analytics component not adequately organized to calculate the various parameters specified in the networks. Hoffmann et al. concentrated on pushing mobile network development towards cost-efficient IT-based solutions utilizing standardized hardware and software-based thoughts such as SDN and NFV [22].

### 4) Cluster Head Selection Approaches

Some previous works provided a different explanation of CHS, but they are not satisfactory enough. For instance, to maximize the reachability of the nodes, Farman et al. proposed a head selection scheme considering the distance and energy [46]. Again, Behera et al. presented an algo- rithm based on residual energy, which is the optimization of LEACH protocols that ensures the adaptation of the procedures [47]. Besides, the nodes that refer to the sensors in IoT were sorted with respect to energy level, and eventually, the authors were able to save some energy [44], [45]. On the contrary, Aslam et al. desired to improve the QoS, and they used a new Integer Linear Program (ILP) to allocate a channel when a CH was selected on a rational basis [55]. For more clarification, Table 4 illustrates the summary with the novelty of the proposed procedure comparing to the previous algorithms.

### 5) Blockchain with SDN

Shao et al. introduced an established consensus algorithm as Simplified Functional Byzantine Fault Tolerance (SPBFT) model for stable [23] message communications to the SDN network. Also, to improve the SDN control layer's safety and consensus technique, a Blockchain is intended to con- tribute to the design of the SDN protection network. Besides, they closely resembled the different tests and evaluated multiple parameters with the BFT algorithm. In a similar kind of research, Chaudhary et al. used two pre-eminent technologies, such as Blockchain and SDN to enhance the network's service quality (QoS) [24]. The authors also introduced BEST, which is Blockchain-based energy trading in the SDN environment. However, they did not consider the various energy sources, which used more effectively in the desired networks. Qiu et al. proposed an imminent unanimity Blockchain-based permit in distributed SDIIoT [25]. Within the suggested scheme, the authors have effectively used a new duelling deep Q-learning method for better simulation operation. The "DistBlockNet" model for IoT architecture [26] as introduced in another work by Sharma et al. This architecture enables networking technology to earn two significant incentives, such as SDN and Blockchain. The authors suggested a method for updating a table of flow rules using a Blockchain methodology as well as formalizing the table of flow rules. The authors have measured the results in distinct metrics in which their work provided, with a better outcome comparing to the surviving work.

### 6) Blockchain for Smart Condominium

Alcarria et al. contributed to Consumer Information management and Collaborative mechanisms in the area of smart Communities, which is needed for the exchange of resources [7]. Authors proposed an authorization system that allows more reliable management and authorization functionalities validated in application scenarios where Blockchain was used as an information exchange platform for transaction privacy and security. Nasarre et al. hashed out different types of collaborative housing and found some potentials of Blockchain technology to alleviate access to housing [12]. They cited the impact of the collaborative economy in accessing, funding, and organizing housing. To make the transaction cost-effective and for security purposes, the authors used the idea of Blockchain. Mital et al. implemented a smart community management and control system based on cloud [2]. The involvements of a condominium controlled by Smart Community, which is very flexible, and it worked with visibility and transparency. They provided a better understanding of the framework of cloud computing-based smart community services, suitable in countries like India. The authors in [1] discussed condominium Boom in the re-urbanization and re-evolution of the inner city in Canada. Authors in their paper highlighted the changes of a rural area due to the deployment of the condominium.

**TABLE 2.** Existing works analysis on Blockchain

| Works | Platform | Contributions | Architecture | Security & Privacy Issues Address | Consensus Protocol |
|---|---|---|---|---|---|
| Alcarria et al. [7] | Smart Communities | An authorization process for resource monitoring and trading for smart communities | Distributed | Yes | Yes |
| Ghandour et al. [49] | Smart City | Blockchain based applications in smart cities | Distributed | Yes | Yes |
| Anich et al. [45] | Smart Building | DistBlockBuilding architecture for smart building using SDN and blockchain technologies | Distributed | Yes | No |
| Sharma et al. [50] | Smart City | SDN and Blockchain based hybrid architecture for smart cities | Distributed | Yes | Yes |
| Gu et al. [51] | Cloud Computing | Decentralized transaction system in cloud storage based on smart contract | Distributed & Decentralized | Yes | No |
| Yazdinejad et al. [52] | IoT Networks | SDN & Blockchain based secure and energy-efficient architecture for IoT networks using clustering structure | Distributed | Yes | Yes |
| Singh et al. [53] | Smart city | Fog and Blockchain model for internet of everything's in smart cities | Distributed & Centralized | Yes | No |
| Sharma et al. [54] | Automotive Industry & Smart city | Distributed Automotive Industry architecture based on Blockchain for Smart cities | Distributed | Yes | Yes |

**TABLE 3.** Comparison of existing works with proposed work

| Works | Blockchain | SDN-IoT | NFV | Multi-Tier Environments | CHS Algorithm |
|---|---|---|---|---|---|
| Ramon et al. [7] | Yes | No | No | No | No |
| Sinh et al. [15] | No | Yes | Yes | Yes | No |
| Islam et al. [20] | Yes | Yes | Yes | Yes | No |
| Abdulqadder et al. [21] | No | Yes | Yes | Yes | No |
| Chaudhary et al. [24] | Yes | Yes | No | No | No |
| Sharma et al. [26] | Yes | Yes | No | Yes | No |
| Mukherjee et al. [31] | No | Yes | Yes | No | No |
| Rahman et al. [32] | Yes | Yes | No | No | No |
| Rahman et al. [44] | Yes | Yes | Yes | No | Yes |
| Anich et al. [45] | Yes | Yes | No | No | Yes |
| Ghandour et al. [49] | Yes | No | No | No | No |
| Sharma et al. [50] | Yes | Yes | No | Yes | No |
| Gu et al. [51] | Yes | No | No | No | No |
| Yazdinejad et al. [52] | Yes | Yes | No | Yes | No |
| Sharma et al. [54] | Yes | No | No | No | No |
| Aslam et al. [55] | No | No | No | No | Yes |
| **Proposed** | Yes | Yes | Yes | Yes | Yes |

Table 3 shows the summary of some most related works comparing with the proposed work. Based on the existing literature survey, we observe that most researchers presented the theoretical work and there is a potential scope to contribute to the technological aspects, e.g., integration of technologies such as IoT, Blockchain, SDN with NFV towards smart applications. The authors propose a conceptual framework of intelligent condominium networking involving SDN and Blockchain technology in this regard. In addition, the authors also present a cluster head selection approach to clump the sensors efficiently and low-power dissipation transmission of data.

**TABLE 4.** Existing works analysis on cluster head selection procedure

| Works | Node Sorting with Energy Values | Euclidean Distance | Optimization | Energy Savings |
|---|---|---|---|---|
| Farman et al. [46] | No | Yes | No | Yes |
| Behera et al. [47] | No | No | Yes | No |
| Rahman et al. [44] | Yes | No | No | Yes |
| Anich et al. [45] | Yes | No | No | Yes |
| Aslam et al. [55] | No | No | Yes | Yes |
| **Proposed** | Yes | Yes | Yes | Yes |

## III. PROPOSED "DISTB-CONDO" ARCHITECTURE WITH NFV FOR SMART CONDOMINIUM

For a smart condominium, a secure distributed block condominium architecture has been proposed which is shown in Fig. 2. The SDN technology is applied in the network to build up a powerful and effective IoT network. Moreover, to analyze the architecture of strong-arm networks and enhance the safety of smart condominiums purposes, this study has also applied Blockchain technology. Afterwards, decentralized and secured Blockchain networking approach is presented for enhancing the data security as well as confidentiality. The pieces of information gathered from the IoT sensors are stocked into the cloud and data centres simultaneously. The possibility of data larceny is very high in the cloud because many users use the cloud at the same time, and it is not quite hard to steal data over the Internet. So as soon as the data entered into the cloud, it will go through the technology of Blockchain and will be stored in block by block with a ledger ensuring the data security.

The authors have divided the whole architecture scenario into several distinct phases. Such as, at first IoT forwards data to provide the facility to extract sensor information, Next, to control the entire network, SDN utilizes SDN-IoT enabled common gateway including data and control layer. On the other hand, NFV technique is used for data virtualization. The whole architecture is shown in Fig. 2 and discussed thoroughly in the following sections.

### A. THE FLOW DIAGRAM OF THE PROPOSED MODEL

The flowchart depicts the proposed model's data flow, starting from the data source, i.e., building IoT devices to the cloud or Distributed Hash Table (DHT) storage which is shown in Fig. 3. Firstly, the SDN-NFVI integrated smart edge gateway controls source data and thus eliminates unwanted data. Secondly, SDK APIs with solidity codes (smart contract for Ethereum) transform data into the transaction (Tx) fashion and lodge it to the Blockchain peers subject to verification. Upon successfully receiving the Tx, the respective protocols, i.e., gossip protocols, run the consensus mechanism. Once Tx gets consented by the associated peers, it updates the ledger, dissipates the latest update through the distributed network. Finally, the Tx is added to the ledger, and the respective data is sent to the storage nodes, i.e., DHT, AWS, etc. for further preservation and maintenance.

### B. IOT FORWARDING DEVICES AND CLUSTER SENSORS HEAD SELECTION PROCEDURE

IoT forwarding devices can forward data through SDN common gateways controller like smart storage devices, lighting, smart TV, etc. SDN dynamic gateway controller and OpenFlow protocol can filter the data received from IoT devices together. Further, this work used "DistB-Condo" to extend the security of data in the employment surface layer in the bright condominium blocks. But the execution of their different functions depends on the success of selecting the cluster head correctly. In this incision, a cluster head selection procedure is shown in Fig. 4.

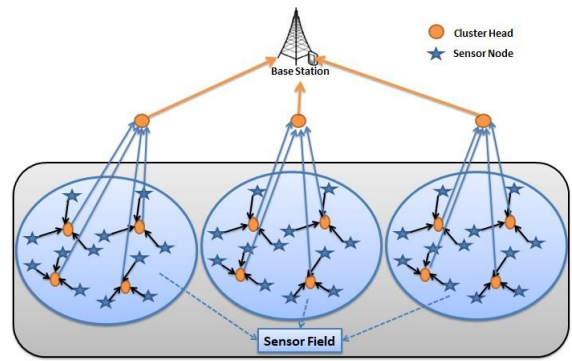

**FIGURE 4.** Clusters Head Selection approach

Thereafter, this article hashed out the implications of the procedural steps to pick cluster head. It also has selected the cluster head (CHs) among different clusters that transmit data over the Internet. The cluster heads (CHs) can forward the data to the base station. Lowering power expenditure and enhancing the lifespan of a network are the main goal of the algorithm. The algorithm pays attention to the initial energy and optimizes the values of cluster heads so that it can choose the heads for the network that is suitable for IoT forwarding devices.

By the utilization of the proposed algorithm,

- The heads for each cluster are selected from the sensors connected within a network.
- S is the set of nodes, and the data structure has several properties such as energy, area which defines area covered by a node, location in 3D plane containing X,Y, and Z, etc.

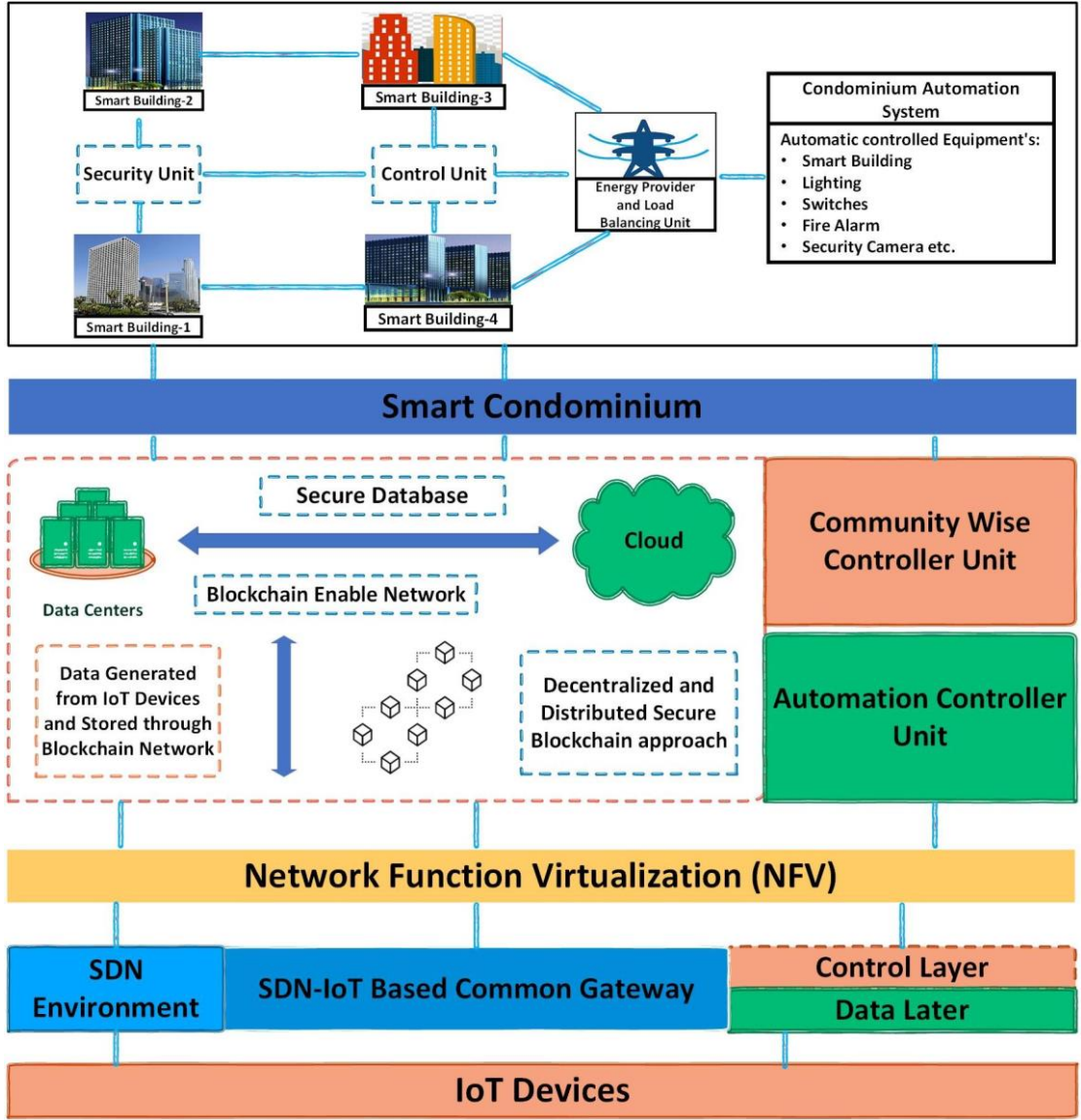

**FIGURE 2.** Proposed architecture of "DistB-Condo"

- Firstly, S is sorted based on the energy of nodes and then the distance from the base station.
- The first node which has the highest energy is selected for being head within a cluster and it is the closest to the Base Station at the same time.
- Then, the members are assigned under the cluster head.
- Euclidean distance from equation 1 is used to assess the distance between two nodes.

$$d(\mathbf{p}, \mathbf{q}) = \overline{(p_x - q_x)^2 + (p_y - q_y)_2 + (p_z - q_z)_2} \quad (1)$$

In the algorithm 1, this study uses head and member as the control variables to notify whether a sensor is a cluster head or a member of another cluster head. Further, CH is returned with the list of cluster heads and its members. Table 5 depicts the parameters of the algorithm.

Table 4 clearly shows the novelty of the proposed cluster head selection algorithm. A good number of researches demonstrates how to choose a cluster head but each of them has some limitations. For example, the article by Farman et al. [46] requires optimization of the parameters, and the nodes were not sorted with the level of energy. The algorithm from [45] did not use the euclidean distance, which works better in reality. On the contrary, the proposed algorithm has the degree of features.

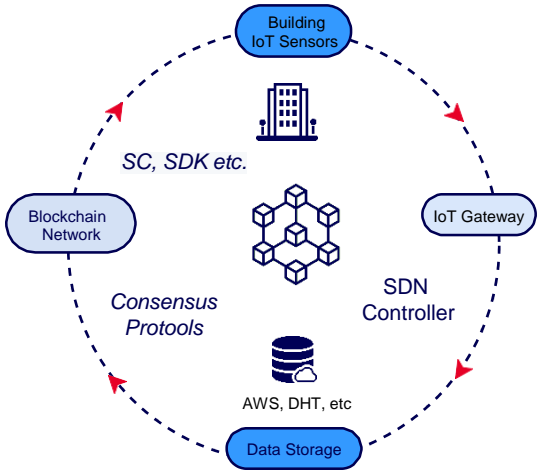

**FIGURE 3.** The flow diagram of the proposed model

### C. SDN ENVIRONMENT WITH INTELLIGENT GATEWAY PROTOCOL

SDN is a networking paradigm that aims to manage and configure the networking application. It can be able to provide some features for networking system such as logically centralized control yet dynamic in nature, network programmability. Further, comparing with traditional networking features, it provides better network programmability and controllability. Also, it can be capable of managing the security threat, new challenges, and different types of attacks than the existing traditional networking model appropriately. It won't control different types of wiretaps and vulnerabilities [33]. This section contains an SDN environment, which includes Data, Control, and Application Layer, as depicted in Fig. 5.

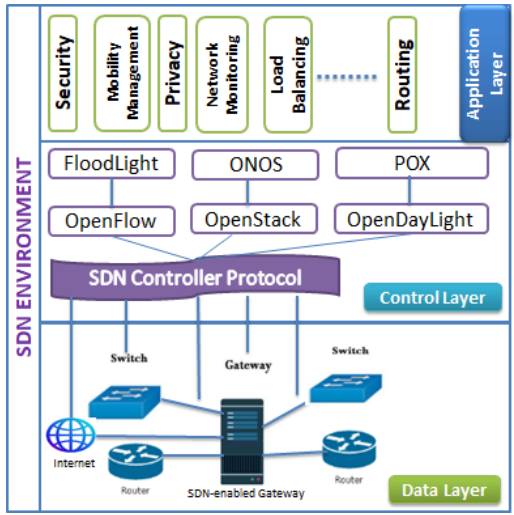

**FIGURE 5.** A conceptional architecture of SDN

#### 1) Data Layer

In SDN, the data layer is also referred to as an infrastructure layer. It is the most bottom layer in the SDN environment,

**Algorithm 1:** Proposed Cluster Head Selection Algorithm

**Input:** n: Number of Nodes, S: Set of nodes
**Output:** CHs: Cluster Heads

```
1  while (1) do
2      for i ← 1 to n − 1 do
3          min = i
4          for j ← i + 1 to n do
5              if (S[j].energy < S[min].energy) then
6                  min = j
7              end
8              swap(S[i], S[min])
9          end
10     end
11     for i ← 1 to n do
12         S[i].distBs ← Edist(BS, S[i])
13     end
14     for i ← 1 to n − 1 do
15         min = i
16         for j ← i + 1 to n do
17             if (S[j].distBS < S[min].distBS) then
18                 min = j
19             end
20             swap(S[i], S[min])
21         end
22     end
23     for i ← 1 to n do
24         if S[i].head = False &
            S[i].member = False then
25             S[i].head ← True
26             for j ← i + 1 to n do
27                 if S[i].area > Edist(S[i], S[j]) then
28                     S[i].members.append(S[j])
29                     S[j].member ← True
30                 end
31             end
32         end
33     end
34     for i ← 1 to n do
35         if S[i].head = True then
36             CH.append(S[i])
37         end
38     end
39 end
```

as depicted in Fig. 5. This plane offers SDN enabled gateway for connecting IoT forwarding devices (router, switch, firewall, storage, etc.) efficiently. It can also provide two type switches like virtual switches related to software-based switches commonly run on Linux operating system. Another one is physical switches that are related to hardware-based switches; it's basically utilizing the higher flow of physical forwarding devices in the infrastructure plane of the SDN architecture. This also switches responsible for forwarding, expending, and exchanging network packets in the network-based applications [34]. Moreover, the network forwarding devices and SDN controllers are communicated with each other by using a more secure TLS connector. Then, the data plane and SDN controller(s) communicate with each other by using the OpenFlow protocol [35]. After that, all secured data is going through the control layer. Furthermore, the data plane captures all IoT forwarding data from a smart condominium environment.

**TABLE 5.** Parameters of CHS Algorithm

| Parameters | Definition |
| --- | --- |
| n | Number of Nodes |
| S | Set of Nodes |
| CHs | Cluster Heads |
| Edist | Euclidean Distance |
| BS | Base Station |
| D (p,q) | Euclidean distance between two nodes |
| Edist (BS, S[i]) | Euclidean distance among Base station and Set of Nodes |

#### 2) Control Layer
The Control Plane (CP) is the main backbone of the SDN architecture. A controller includes components as primary such as a logical central and functional controller. The logic controller provides a high facility for network communication. Also, CP can be capable of mapping between the forwarding and applications layer in the SDN architecture. It offers different types of networking services for the desired platform [34]. Furthermore, the control plane architecture conveys some tremendous protocols such as NOX, OpenFlow, POX, Floodlight, OpenDayLight, Openstack, and Beacon [36]. After that, the controller introduced some interfaces named southbound, northbound, and eastbound interfaces to interact appropriately. Also, this controller enhances the networking system, which is able to utilize the high security and privacy of data in the smart condominium architecture.

#### 3) Application Layer
The topmost layers are called the application layer in the SDN architecture. Then, the SDN-based scheme has committed a huge number of tending for dynamic updating of forwarding flow rules efficiently. Also, the application plane enhances the networking services between control and application platform over the physical forwarding objects or virtual objects. After that, it admits more prominent stages of network configuration and management named network data analytics, or specialized functions expecting to treat in large data centres [37]. This layer allows several services like smart computations, security, smart optimization, mobility management, load balancing, routing computation, switching, reliability, as well as network monitoring in the smart condominium network, as shown in Fig. 5.

#### 4) Topology Design
Fig. 6 displays the topology of 50 network nodes. For network topology with 50 nodes, it can be considered as 9 access points (APs), with a total number of 46 stations (sta) connected to them. It can start communication between nodes after designing the topology by pinging them and capturing the traffic flow packets for each topology in the Wire shark platform. Besides, the graphs can be carefully analyzed for this network to make decisions about the performance and effectiveness of each topology.

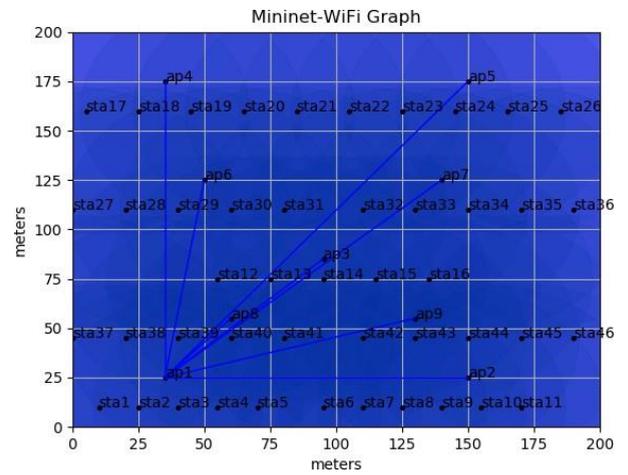

**FIGURE 6.** Network Topology with 50 nodes

### D. ROLE OF NETWORK FUNCTION VIRTUALIZATION (NFV)
The NFV is the chief orchestrator and collaborator in SDN architecture. Further, NFV provides a virtual platform for SDN-IoT enabled physical environment. Moreover, on the necessity of cloud services, NFV offers quick network packet processing in the 5G network architecture [38]. It can also be able to provide the virtual environment named Virtual Machine (VM) for running a substantial computational operation to the networking paradigm. Furthermore, NFV used for separating the data plane and control plane in the SDN model efficiently. In this case, NFV has been used in this research to virtualize all secured forwarding data, which got from the sensor data level.

#### 1) Load Balancing
NFV provides a VM environment for physical forwarding devices. A load balancing strategy is performed by some efficient components such as a physical server, Hypervisor, and guest virtual machine [39]. Where the physical server offers CPU, storage, and RAM to the network model. Then,

Hypervisor monitors of VM and guest virtual machine utilizing emulator based software and practicalities of a solid platform based on the desired application.

2) Cost Minimization

To reduce the cost of this proposed architecture for smart condominiums, the NFV manager is executed efficiently. NFV manager uses REST API of Open Baton's Network Functions Virtualisation Orchestrator (NFVO), TOSCA CSAR network service, and Network Service Record (NSR) [40]. Also, NFV based application layer used as a cost-effective manner in the desired network mode.

### E. DECENTRALIZED AND SECURED BLOCKCHAIN PROCEDURE

To control the presented "DistB-Condo" architecture, this article proposed an excellent distributed approach named Blockchain. After the operation of SDN-IoT, Blockchain operation is being performed in the system architecture. Basically, this work considers the Blockchain approach in the proposed architecture to ensure high-security. Also, Blockchain can be performed in a distributed manner that facilies such as transparency, confidentiality, high safety, reliability for safe and reliable communication in the network [41]. A distributed Blockchain is a particular type of distributed ledger, which connects one block to another block using a unique chain called Hash chain. In a distributed environment, the first genesis block can be created. It contains some block, timestamp, hash data, and transaction fields.

Fig. 7 points the necessary procedure of the Blockchain for a smart condominium. In intelligent building condominiums, the information collecting devices, i.e., sensors, are connected to the internet via LAN, WiFi, etc. After perceiving the information, these sensors send the data to the SDN Edge Gateway. The data handling process under the SDN protocol has been described already in a previous section. Now, in the second step, as shown in figure 7, the SDN platform forwards the data as a transaction to the Blockchain. The whole Distributed Blockchain consists of several components such as a public ledger (where all the transactions are added as blocks), Consensus Protocols (i.e., distributed proof of stack), Smart Contracts, and Software Development Kit (SDK) to compile and run the functions like verify() and valid(). The operational procedures of transaction processing by the smart contracts will be discussed later with a suitable example. After modifying the public ledger with a new transaction, the data will be stored in Cloud Storage (i.e., AWS, Azure, etc.). It is done with REST or RPC call following some API and protocols, and the Solidity Operations are run by the SDK.

### F. BLOCKCHAIN-BASED TRANSACTION PROCESS IN CONDOMINIUM ENVIRONMENT

This segment provides an idea to the process of a Blockchain transaction, as shown in Fig. 8. In each transaction, there are several overheads along with the data like–the sensor id is the identification of sensor from which the data is coming, the destination field contains the address of the receiver and some other extra information (as shown in the Fig. 8 with a JSON format of the transaction packet).

To validate the transaction and perform the operation with notifying all the other blocks, there is a Smart Contracts into the system that will receive the transaction and check for the validation process based on some proof of work. The transaction will be appended in the list which has already been created. When everything is alright then, the Smart Contact forwards the Tx packet to the Tx processing method, and it checks for the further status of the transaction and broadcasts the new data to its destination and commits some changes across the public ledger, notifying the completion of the transaction.

On the contrary, if any error occurs during the validation process, or if the transaction request contains tempered information, then no changes will be committed to Blockchain and the transaction will be sent to the waiting room. It will be in pending for a certain period, if further identification is not found from the sender, the transaction will be discarded from the system queue.

### G. SMART CONDOMINIUM SERVICES

After performing all technical operations on the IoT sensors data, these data transform one stage to another stage according to the presented "DistB-Condo" architecture. After ensuring that the data is secured, then these data are transformed in smart condominium area for different types of services performed by human, as shown in Fig. 9.

Furthermore, the smart condominium contains some efficient domain named home, community, and service domain [42]. Moreover, the main component of a smart condominium is a sensor database, which can be capable of receiving current capturing data from the IoT sensor network continuously [10]. In addition, this intelligent condominium can be able to reduce noise, traffic congestion, pollution, and so on. Also, it can increase public safety, house security, healthcare caliber, exigency response powers as well as transfer speed of data [43].

## IV. RESULT ANALYSIS AND DISCUSSION
### A. SETTING UP THE ENVIRONMENT

To the setting's environment of the suggested model, this study used Emulator Mininet-Wifi, OpenFlow based SDN rules in the SDN context for giving rise to the goal outcomes. This also used the Ubuntu (Linux) operating system, Core(TM)-i5 processor, 2.20GHz CPU, 6GB RAM, 1 TB ROM, and other external memory for estimating the expected outputs. Additionally, for conceiving the execution of the IoT-SDN network, this experiment utilized the Wireshark platform accurately. Modeling parameters are illustrated in Table 6.

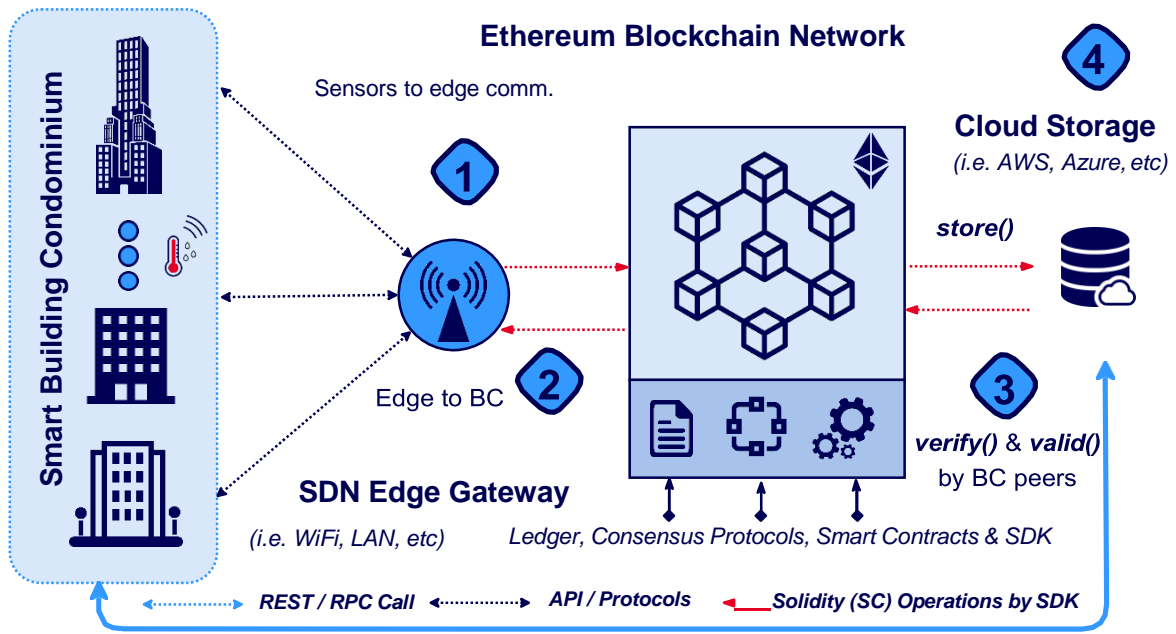

**FIGURE 7.** Distributed Blockchain approaches for smart condominium

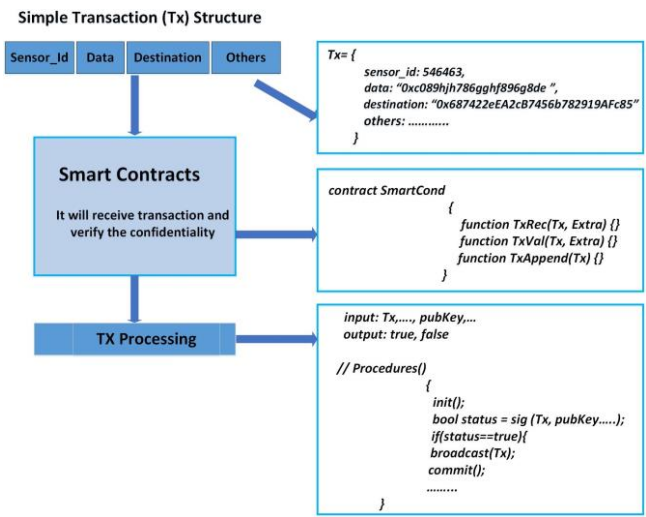

**FIGURE 8.** Blockchain transaction processing

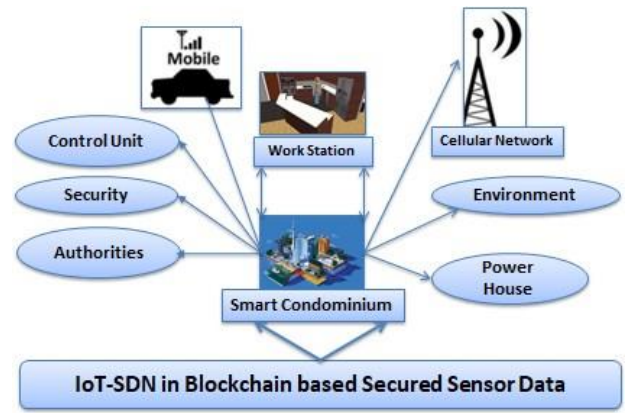

**FIGURE 9.** Overview of smart condominium services

### B. PERFORMANCE EVALUATION OF OPENFLOW-BASED SDN IN IOT NETWORKS

In this section, the performance of the proposed model is evaluated and presented based on key performance indicators. Here, we have used Throughput, packet arrival rate, and file transferring operation properties for measuring the performance.

Firstly, it notices the throughput on the basis of the number of nodes, as shown in Fig. 10. Then, it can be able to show the throughput comparisons between OpenFlow based SDN and the proposed model "DistB-Condo" practically. Moreover, this article also noticed that when the number of nodes is less than 3, the throughput is almost similar or no difference. But after completing a certain period of time, when the number of nodes is increasing, the throughput is also increased. For 50 nodes, throughputs are close to each other, but after increasing the nodes throughput also increased.

Secondly, the packet arrival rate also visualizes the performance analysis between the proposed system vs OpenFlow based SDN, as depicted in Fig. 11 properly.

At the same time, it can be capable of showing the packet arrival rate (thousand/s) comparisons with bandwidth (Mb/s). Then, it is notified that the bandwidth starts at point 3.8 (approximately) for both OpenFlow based SDN and distributed Blockchain-based SDN (proposed "DistB-Condo"). After executing a little period, with increasing the packet arrival rate, the bandwidth is gradually decreased based on increasing the DDoS attack rate. On the contrary, the performance of the presented model slightly affected by increasing the attack rate. In the proposed model, the drop of the bandwidth is 11% in the presence of the DDoS attack. Finally, when the packet arrival rate reached 32, the bandwidth is increased for DistB-Condo. However, bandwidth for the OF-Based SDN started to go down and the drop of the bandwidth is 81% in the presence of DDoS attack.

TABLE 6. Simulation Setup

| Parameters Name | Values |
|---|---|
| *General Parameters* | |
| Network emulator | Mininet-WiFi |
| Packet analyzer | Wireshark |
| Programming language | Python |
| *SDN Parameters* | |
| SDN routing protocol | OpenFlow |
| Number of SDN controllers | 5 |
| No. of Gateways | 2 |
| *Blockchain Parameters* | |
| Blockchain platform | Ethereum |
| Consensus protocol | Proof of Work (PoW) & Proof of Stake (PoS) |
| *Other Parameters* | |
| Number of IoT devices | $1 - 100$ |
| Simulation time | $500\ s$ |
| Simulation area | $2500\ m \times 2500\ m$ |
| Data rate | $10\ Mbps$ |
| Transmitted packet size | $128 - 1024\ bytes$ |

TABLE 7. Number of Iot nodes are executed with respect to throughput

| | Throughput (kbps) | |
|---|---|---|
| No. of Nodes | DistB-Condo | OF- Based OpenFlow |
| 1 | 2 | 2 |
| 5 | 4 | 3 |
| 10 | 8 | 7 |
| 15 | 11 | 8 |
| 20 | 13 | 9 |
| 25 | 16 | 12 |
| 30 | 18 | 15 |
| 35 | 20 | 17 |
| 40 | 23 | 19 |
| 45 | 25 | 20 |
| 50 | 26 | 23 |
| 55 | 27 | 24 |
| 60 | 29 | 25 |

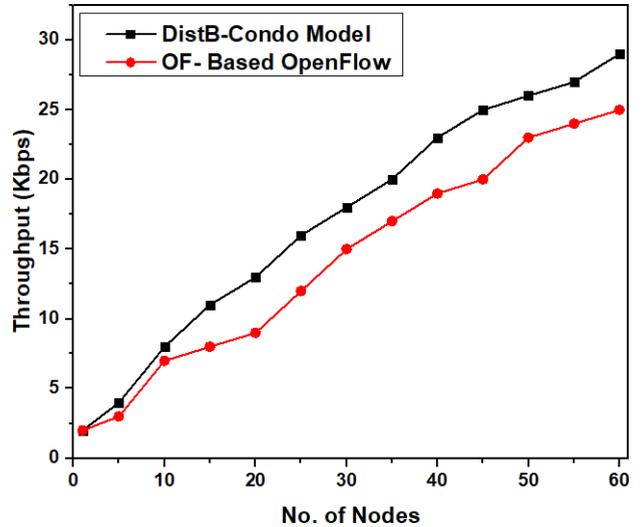

FIGURE 10. Comparison of throughput when changing no. of IoT devices of the proposed model with OF-Based SDN

TABLE 8. Packet arrival rate with respect to Bandwidth

| | Bandwidth (Mbps) | |
|---|---|---|
| Packet Arrival Rate | DistB-Condo | OF- Based OpenFlow |
| 1 | 3.8 | 3.7 |
| 4 | 3.7 | 2.8 |
| 7 | 3.7 | 2.7 |
| 10 | 3.6 | 2.1 |
| 13 | 3.6 | 1.9 |
| 16 | 3.5 | 1.8 |
| 19 | 3.5 | 1.5 |
| 22 | 3.4 | 1.3 |
| 24 | 3.4 | 1.2 |
| 28 | 3.3 | 0.9 |
| 32 | 3.4 | 0.3 |

Again, Fig. 12 shows the file operation based on the core and proposed model. This graph demonstrates the file operation on the basis of the response time and size of the file efficiently. With the increasing number of file sizes, response time is also rising, and it performed better in the system. For less number of file transfer, response time is slightly changed. But, after increasing the file transfer, the response time started to go high. Furthermore, this work noticed; the presented model can be able to transfer a substantial file than the actual core based system. Finally, the result shows that the proposed model 5% less response time on average compared to the core model.

## C. PERFORMANCE EVALUATION OF PROPOSED MODEL USING ETHEREUM IN IOT NETWORKS

Fig. 13 shows gas consumption for no. of transactions. Gas limit represents that for a single transaction, what amount of energy (maximum) is required. For small no. of operations, the gas limits are very negligible. But when we are increased the no. of the transaction, the gas limit is also increased. Up to 15 transactions graph changing very high. Then, the graph is growing linearly when the no. of transactions increased.

TABLE 9. Packet arrival rate with respect to Response time(ms)

| | Response Time(ms) | |
|---|---|---|
| File Sizes(Mb) | DistB-Condo | Core model |
| 2 | 85 | 140 |
| 4 | 290 | 430 |
| 8 | 460 | 560 |
| 16 | 690 | 780 |
| 32 | 855 | 990 |
| 64 | 1190 | 1290 |
| 128 | 1510 | 1620 |
| 256 | 1700 | 1810 |
| 512 | 1780 | 1880 |
| 1024 | 1930 | 2020 |

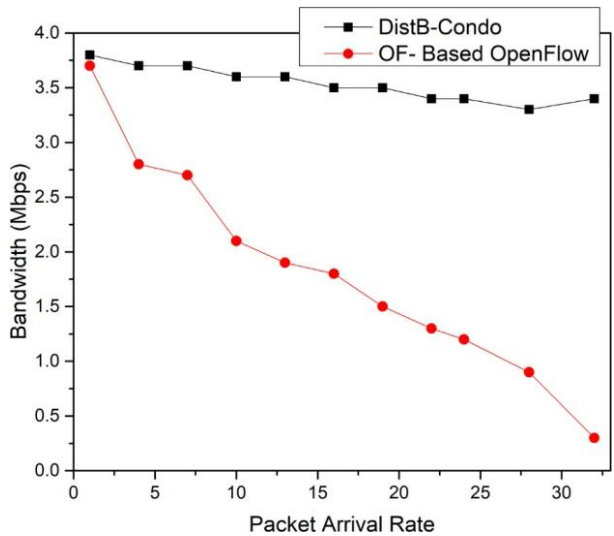

**FIGURE 11.** Comparison of bandwidth of the proposed model with respect to OF-Based SDN in the presence of attack

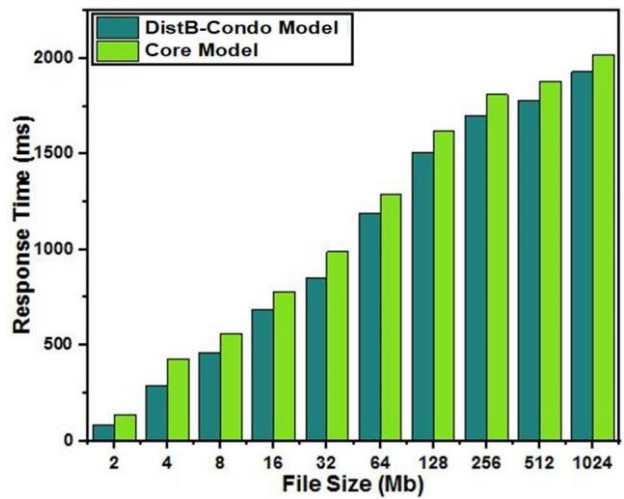

**FIGURE 12.** Comparison of response time(ms) with respect to transfer of files(Mb) of the proposed architecture with core model

### D. PERFORMANCE ANALYSIS DURING ATTACKS

After performing the comparisons among throughput, packet arrival rate analysis, file operation, and RTT analysis, it is evident that the proposed "DistB-Condo" architecture has shown better performances compare with OpenFlow based SDN model.

In this segment, this experiment also observed a Distributed Denial-of-Service (DDoS) attack in the implementation area accurately. Fig. 14 shows the analysis of CPU utilization for DDoS attack in experimental environment during various applications are running continuously. Further, there also used learning set for recording CPU utilization during a Flooding attack. Moreover, it also shows average CPU utilization in different applications based on "DistB-Condo" scheme, when DDoS based Flooding attacks are performed. Then, this attack started at point 2.9 value approximately, initially, the attack is performed a normal process in the system. But with increasing the time the attack rate is also increasing. In proceeding with the attack rate after a specific time, this work observed that "DistB-Condo" provides sufficient protection against this attack efficiently. Furthermore, Table 12 shows the average CPU utilization for 12 & 13 learning application.

**TABLE 10.** Gas consumption with respect to no. of transactions of the proposed model

| No. of transactions | Gas Consumption |
|---|---|
| 3 | 25000 |
| 6 | 34000 |
| 9 | 44000 |
| 12 | 53000 |
| 15 | 64000 |
| 18 | 74000 |
| 21 | 84000 |
| 24 | 95000 |

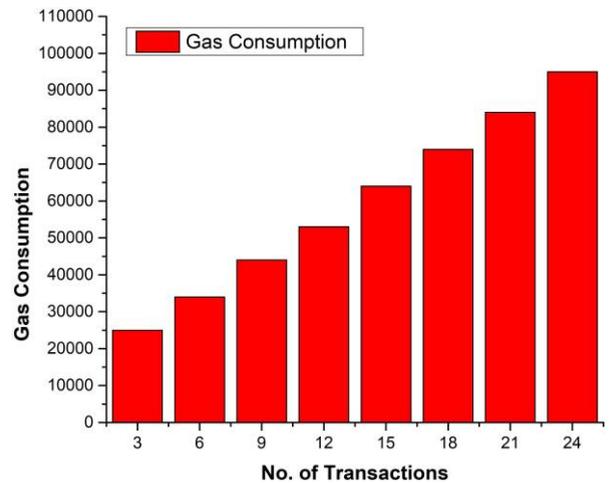

**FIGURE 13.** Gas consumption with respect to no. of transactions of the proposed model

**TABLE 11.** CPU Utilization during Flooding attack

| Time(s) | CPU Utilization(%) |
|---|---|
| 0.2 | 3 |
| 0.4 | 3 |
| 0.6 | 7 |
| 0.8 | 9 |
| 1.0 | 16 |
| 1.2 | 15 |
| 1.4 | 21 |
| 1.6 | 27 |
| 1.8 | 25 |
| 2.0 | 23 |
| 2.2 | 21 |
| 2.4 | 18 |
| 2.6 | 11 |
| 2.8 | 10 |
| 3.0 | 8 |
| 3.2 | 2 |

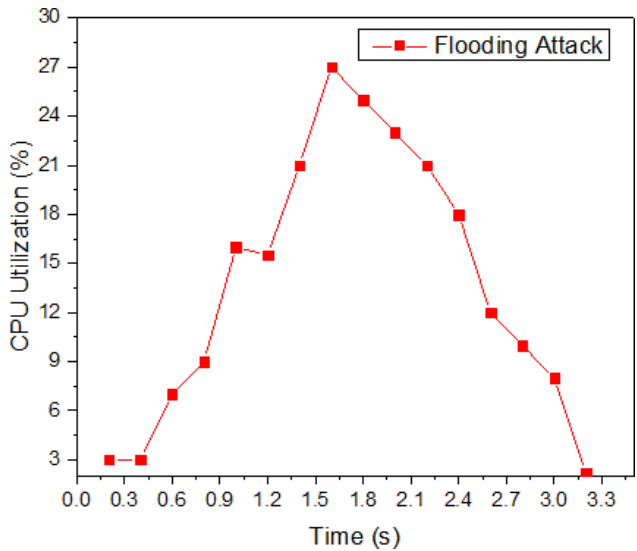

FIGURE 14. CPU Utilization during Flooding attack of the proposed architecture

TABLE 12. Attack comparisons of present work with proposed work

| | CPU Utilization (%) | | | |
|---|---|---|---|---|
| | 12 Learning apps | | 13 Learning apps | |
| Time (s) | Sharma et al. [26] | Proposed | Sharma et al. [26] | Proposed |
| 0.5 | 0.5 | 0.4 | 0.2 | 0.2 |
| 0.6 | 2.1 | 1.7 | 0.5 | 0.4 |
| 0.8 | 13.3 | 9.6 | 15.2 | 13.4 |
| 1.0 | 6.7 | 4.6 | 8.3 | 6.4 |
| 1.2 | 7.8 | 6.1 | 7.7 | 5.4 |
| 1.6 | 2.2 | 1.6 | 0.3 | 0.3 |
| 2.0 | 0.4 | 0.3 | 0.1 | 0.1 |

## V. FUTURE SCOPES
### A. MACHINE LEARNING
Nevertheless, as time is going on, more novel problems are revealing; that's why such kinds of technologies need more comprehensive research. IoT generates a massive number of data and transmits those to the controllers, sometimes the information can be corrupted, and to take the decision, the system may have some artificially intelligent agents [56]. In this phenomenon, Machine Learning can play a vital role.

### B. SECURITY AND PRIVACY
The volume of data that will be generated within the system is not imaginable, so the security and privacy of the information is a great point to remember at every step of the processing [57]. Recently researchers can work out to make the security stricter, and besides SDN, NFV, which is another issuing technology, could be combined for further development.

### C. STORAGE AND DATA TRANSMISSION
A large number of databases are needed to store the data of the smart condominium. At the same time, the storage should be reliable and flexible so that the data could be transmitted efficiently to the destination [58]. It is always a real challenge to save the information all together and provide access to the stora.3.2ge publicly in real-time with proper stability.

### D. IMPLEMENTATION OF BLOCKCHAIN
From another point of view, researchers can keep the implementation of Blockchain on their eye as the effectuation of this technology is not satisfactory and confusing to the end-users. Applying Blockchain in the real world could be critical to the researchers [59].

In the future, more parameters such as time scaling, resource sharing as well as quick response time based on numerous requests could be added effectively to the presented architecture.

## VI. CONCLUSION
In this study, we have proposed a distributed secure Blockchain-based SDN-IoT model for the smart condominiums. We present a CHS algorithm that selects CH with the highest energy in an optimal way. In addition, the SDN controller is used to continuously monitors and manages the IoT devices data in the entire IoT networks; it also detects the possible attack in the networking system; increases the scalability, flexibility issues. Then, NFV provides a virtual platform for the SDN-IoT-enabled physical environment and also saving energy, enhancing the lifetime of the whole network. Moreover, distributed Blockchain provides adequate security & privacy; it detects and mitigates cyber-attacks in the proposed system efficiently. To evaluate the performance of the proposed model various parameters (e.g. throughput, packet arrival rate, response time, and gas consumption) are used; the outcome of our assessment shows higher throughput, higher bandwidth in the presence of attacks, and takes less response time with respect to the file size. Finally, we have compared our proposed architecture with two existing systems where the proposed model performs better than others in all aspects. Furthermore, we have some limitations including we are not considered an end-to-end delay, packet delivery ratio, some other attacks that are closely related to IoT ecosystems as well as smart condominiums.

## REFERENCES
[1] U. Lehrer, R. Keil, and S. Kipfer, "Reurbanization in Toronto: Condominium Boom and Social Housing Revitalization," disP-The Planning Review, vol. 46, no. 180, pp. 81–90, 2010.
[2] M. Mital, A. K. Pani, S. Damodaran, and R. Ramesh, "Cloud based Management and Control System for Smart Communities: A Practical Case Study," Computers in Industry, vol. 74, pp. 162–172, 2015.
[3] J. P. Martins, J. C. Ferreira, V. Monteiro, J. A. Afonso, and J. L. Afonso, "IoT and Blockchain Paradigms for EV Charging System," Energies, vol. 12, no. 15, p. 2987, 2019.

[4] A. P. Plageras, K. E. Psannis, C. Stergiou, H. Wang, and B. B. Gupta, "Efficient IoT-based Sensor Big Data Collection–Processing and Analysis in Smart Buildings," Future Generation Computer Systems, vol. 82, pp. 349–357, 2018.

[5] S. Chakrabarty and D. W. Engels, "A Secure IoT Architecture for Smart Cities," in 2016 13th IEEE annual consumer communications & networking conference (CCNC). IEEE, 2016, pp. 812–813.

[6] P. Bull, R. Austin, E. Popov, M. Sharma, and R. Watson, "Flow based Security for IoT Devices Using an SDN Gateway," in 2016 IEEE 4th International Conference on Future Internet of Things and Cloud (FiCloud). IEEE, 2016, pp. 157–163.

[7] R. Alcarria, B. Bordel, T. Robles, D. Martín, and M.-Á. Manso-Callejo, "A Blockchain-based Authorization System for Trustworthy Resource Monitoring and Trading in Smart Communities," Sensors, vol. 18, no. 10, p. 3561, 2018.

[8] M. Ojo, D. Adami, and S. Giordano, "A SDN-IoT Architecture with NFV Implementation," in 2016 IEEE Globecom Workshops (GC Wkshps). IEEE, 2016, pp. 1–6.

[9] K. Almustafa and M. Alenezi, "Cost Analysis of SDN/NFV Architecture over 4G Infrastructure," Procedia computer science, vol. 113, pp. 130–137, 2017.

[10] N. M. Boers, D. Chodos, J. Huang, P. Gburzynski, I. Nikolaidis, and E. Stroulia, "The Smart Condo: Visualizing Independent Living Environments in a Virtual World," in 2009 3rd International Conference on Pervasive Computing Technologies for Healthcare. IEEE, 2009, pp. 1–8.

[11] S. El Bouanani, O. Achbarou, M. A. Kiram, and A. Outchakoucht, "Towards Understanding Internet of Things Security and Its Empirical Vulnerabilities: A Survey."

[12] S. Nasarre-Aznar, "Collaborative Housing and Blockchain," Administration, vol. 66, no. 2, pp. 59–82, 2018.

[13] A. Kaur, A. Nayyar, and P. Singh, "Blockchain: A path to the future," Cryptocurrencies and Blockchain Technology Applications, pp. 25–42, 2020.

[14] M. U. Younus, S. ul Islam, I. Ali, S. Khan, and M. K. Khan, "A Survey on Software Defined Networking Enabled Smart Buildings: Architecture, Challenges and Use Cases," Journal of Network and Computer Applications, 2019.

[15] D. Sinh, L.-V. Le, B.-S. P. Lin, and L.-P. Tung, "SDN/NFV— A New Approach of Deploying Network Infrastructure for IoT," in 2018 27th Wireless and Optical Communication Conference (WOCC). IEEE, 2018, pp. 1–5.

[16] S. N. Matheu, A. Robles Enciso, A. Molina Zarca, D. Garcia-Carrillo, J. L. Hernández-Ramos, J. Bernal Bernabe, and A. F. Skarmeta, "Security architecture for defining and enforcing security profiles in dlt/sdn-based iot systems," Sensors, vol. 20, no. 7, p. 1882, 2020.

[17] M. Conti, P. Kaliyar, and C. Lal, "Censor: Cloud-enabled secure iot architecture over sdn paradigm," Concurrency and Computation: Practice and Experience, vol. 31, no. 8, p. e4978, 2019.

[18] K. Kalkan and S. Zeadally, "Securing Internet of Things with Software Defined Networking," IEEE Communications Magazine, vol. 56, no. 9, pp. 186–192, 2017.

[19] A. A. Barakabitze, A. Ahmad, R. Mijumbi, and A. Hines, "5g network slicing using sdn and nfv: A survey of taxonomy, architectures and future challenges," Computer Networks, vol. 167, p. 106984, 2020.

[20] M. J. Islam, M. Mahin, S. Roy, B. C. Debnath, and A. Khatun, "DistBlackNet: A Distributed Secure Black SDN-IoT Architecture with NFV Implementation for Smart Cities," in 2019 International Conference on Electrical, Computer and Communication Engineering (ECCE). IEEE, 2019, pp. 1–6.

[21] I. Abdulqadder, D. Zou, I. Aziz, B. Yuan, and W. Dai, "Deployment of robust security scheme in sdn based 5g network over nfv enabled cloud environment," IEEE Transactions on Emerging Topics in Computing, 2018.

[22] M. Hoffmann, M. Jarschel, R. Pries, P. Schneider, A. Jukan, W. Bziuk, S. Gebert, T. Zinner, and P. Tran-Gia, "SDN and NFV as Enabler for the Distributed Network Cloud," Mobile Networks and Applications, vol. 23, no. 3, pp. 521–528, 2018.

[23] Z. Shao, X. Zhu, A. M. Chikuvanyanga, and H. Zhu, "Blockchain-based sdn security guaranteeing algorithm and analysis model," in International Conference on Wireless and Satellite Systems. Springer, 2019, pp. 348–362.

[24] R. Chaudhary, A. Jindal, G. S. Aujla, S. Aggarwal, N. Kumar, and K.-K. R. Choo, "Best: Blockchain-based secure energy trading in sdn-enabled intelligent transportation system," Computers & Security, vol. 85, pp. 288–299, 2019.

[25] C. Qiu, F. R. Yu, F. Xu, H. Yao, and C. Zhao, "Permissioned Blockchain-based Bistributed Software-Defined Industrial Internet of Things," in 2018 IEEE Globecom Workshops (GC Wkshps). IEEE, 2018, pp. 1–7.

[26] P. K. Sharma, S. Singh, Y.-S. Jeong, and J. H. Park, "Distblocknet: A Distributed Blockchains-based Secure SDN Architecture for IoT Networks," IEEE Communications Magazine, vol. 55, no. 9, pp. 78–85, 2017.

[27] H. C. Punjabi, S. Agarwal, V. Khithani, V. Muddaliar, and M. Vasmatkar, "Smart Farming Using IoT," International Journal of Electronics and Communication Engineering and Technology, vol. 8, no. 1, 2017.

[28] D. Kreutz, F. M. Ramos, P. E. Verissimo, C. E. Rothenberg, S. Azodolmolky, and S. Uhlig, "Software-Defined Networking: A Comprehensive Survey," Proceedings of the IEEE, vol. 103, no. 1, pp. 14–76, 2014.

[29] D. Burkhardt, M. Werling, and H. Lasi, "Distributed Ledger," in 2018 IEEE international conference on engineering, technology and innovation (ICE/ITMC). IEEE, 2018, pp. 1–9.

[30] W. Gao, W. G. Hatcher, and W. Yu, "A Survey of Blockchain: Techniques, Applications, and Challenges," in 2018 27th International Conference on Computer Communication and Networks (ICCCN). IEEE, 2018, pp. 1–11.

[31] B. K. Mukherjee, M. S. I. Pappu, M. J. Islam, and U. K. Acharjee, "An SDN based Distributed IoT Network with NFV Implementation for Smart Cities," In press: 2nd International Conference on Cyber Security and Computer Science (ICONCS-2020), Springer, pp. 539–552, 2020.

[32] A. Rahman, U. Sara, D. Kundu, S. Islam, M. J. Islam, M. Hasan, Z. Rahman, and M. K. Nasir, "Distb-sdoindustry: Enhancing security in industry 4.0 services based on distributed blockchain through software defined networking-iot enabled architecture," International Journal of Advanced Computer Science and Applications, vol. 11, no. 9, 2020.

[33] S. W. Pritchard, G. P. Hancke, and A. M. Abu-Mahfouz, "Security in Software-Defined Wireless Sensor Networks: Threats, Challenges and Potential Solutions," in 2017 IEEE 15th International Conference on Industrial Informatics (INDIN). IEEE, 2017, pp. 168–173.

[34] J. Xie, F. R. Yu, T. Huang, R. Xie, J. Liu, C. Wang, and Y. Liu, "A Survey of Machine Learning Techniques Applied to Software Defined Networking (sdn): Research Issues and Challenges," IEEE Communications Surveys & Tutorials, vol. 21, no. 1, pp. 393–430, 2018.

[35] M. Karakus and A. Durresi, "A Survey: Control Plane Scalability Issues and Approaches in Software-Defined Networking (SDN)," Computer Networks, vol. 112, pp. 279–293, 2017.

[36] J. Medved, R. Varga, A. Tkacik, and K. Gray, "Opendaylight: Towards a Modeldriven SDN Controller Architecture," in Proceeding of IEEE International Symposium on a World of Wireless, Mobile and Multimedia Networks 2014. IEEE, 2014, pp. 1–6.

[37] A. S. Thyagaturu, A. Mercian, M. P. McGarry, M. Reisslein, and W. Kellerer, "Software Defined Optical Networks (SDONs): A Comprehensive Survey," IEEE Communications Surveys & Tutorials, vol. 18, no. 4, pp. 2738–2786, 2016.

[38] S. Palkar, C. Lan, S. Han, K. Jang, A. Panda, S. Ratnasamy, L. Rizzo, and S. Shenker, "E2: A Framework for NFV Applications," in Proceedings of the 25th Symposium on Operating Systems Principles, 2015, pp. 121–136.

[39] H. Hawilo, A. Shami, M. Mirahmadi, and R. Asal, "NFV: State of the Art, Challenges, and Implementation in Next Generation Mobile Networks (vEPC)," IEEE Network, vol. 28, no. 6, pp. 18–26, 2014.

[40] S. Vural, R. Minerva, G. A. Carella, A. M. Medhat, L. Tomasini, S. Pizzimenti, B. Riemer, and U. Stravato, "Performance Measurements of Network Service Deployment on a Federated and Orchestrated Virtualisation Platform for 5G Experimentation," in 2018 IEEE Conference on Network Function Virtualization and Software Defined Networks (NFV-SDN). IEEE, 2018, pp. 1–6.

[41] S. Roy, M. Ashaduzzaman, M. Hassan, and A. R. Chowdhury, "Blockchain for IoT Security and Management: Current Prospects, Challenges and Future Directions," in 2018 5th International Conference on Networking, Systems and Security (NSysS). IEEE, 2018, pp. 1–9.

[42] X. Li, R. Lu, X. Liang, X. Shen, J. Chen, and X. Lin, "Smart Community: An Internet of Things Application," IEEE Communications magazine, vol. 49, no. 11, pp. 68–75, 2011.

[43] C. Benevolo, R. P. Dameri, and B. D'Auria, "Smart Mobility in Smart City," in Empowering Organizations. Springer, 2016, pp. 13–28.

[44] A. Rahman, M. J. Islam, F. A. Sunny, and M. K. Nasir, "DistBlockSDN: A Distributed Secure Blockchain based SDN-IoT Architecture with NFV Implementation for Smart Cities," In Press: International Conference on

Innovation in Engineering and Technology (ICIET), vol. 23, p. 24, IEEE, 2019.
[45] A. Rahman, M. K. Nasir, Z. Rahman, A. Mosavi, S. Shahab, and B. Minaei-Bidgoli, "Distblockbuilding: A distributed blockchain-based sdn-iot network for smart building management," IEEE Access, 2020.
[46] H. Farman, B. Jan, H. Javed, N. Ahmad, J. Iqbal, M. Arshad, and S. Ali, "Multi-criteria based zone head selection in internet of things based wireless sensor networks," Future Generation Computer Systems, vol. 87, pp. 364–371, 2018.
[47] T. M. Behera, S. K. Mohapatra, U. C. Samal, M. S. Khan, M. Daneshmand, and A. H. Gandomi, "Residual energy-based cluster-head selection in wsns for iot application," IEEE Internet of Things Journal, vol. 6, no. 3, pp. 5132–5139, 2019.
[48] Z. Qin, G. Denker, C. Giannelli, P. Bellavista, and N. Venkatasubramanian, "A Software Defined Networking Architecture for the Internet-of-Things," in 2014 IEEE network operations and management symposium (NOMS). IEEE, 2014, pp. 1–9.
[49] A. G. Ghandour, M. Elhoseny, and A. E. Hassanien, "Blockchains for smart cities: a survey," in Security in Smart Cities: Models, Applications, and Challenges. Springer, 2019, pp. 193–210.
[50] P. K. Sharma and J. H. Park, "Blockchain based hybrid network architecture for the smart city," Future Generation Computer Systems, vol. 86, pp. 650–655, 2018.
[51] Y. Gu, D. Hou, X. Wu, J. Tao, and Y. Zhang, "Decentralized transaction mechanism based on smart contract in distributed data storage," Information, vol. 9, no. 11, p. 286, 2018.
[52] A. Yazdinejad, R. M. Parizi, A. Dehghantanha, Q. Zhang, and K.-K. R. Choo, "An energy-efficient sdn controller architecture for iot networks with blockchain-based security," IEEE Transactions on Services Computing, 2020.
[53] P. Singh, A. Nayyar, A. Kaur, and U. Ghosh, "Blockchain and fog based architecture for internet of everything in smart cities," Future Internet, vol. 12, no. 4, p. 61, 2020.
[54] P. K. Sharma, N. Kumar, and J. H. Park, "Blockchain-based distributed framework for automotive industry in a smart city," IEEE Transactions on Industrial Informatics, vol. 15, no. 7, pp. 4197–4205, 2018.
[55] S. Aslam, N. U. Hasan, J. W. Jang, and K.-G. Lee, "Optimized energy harvesting, cluster-head selection and channel allocation for iots in smart cities," Sensors, vol. 16, no. 12, p. 2046, 2016.
[56] T. Marwala and B. Xing, "Blockchain and artificial intelligence," arXiv preprint arXiv:1802.04451, 2018.
[57] , Khan, Fakhri Alam and Asif, Muhammad and Ahmad, Awais and Alharbi, Mafawez and Aljuaid, Hanan, "Blockchain technology, improvement suggestions, security challenges on smart grid and its application in healthcare for sustainable development", Sustainable Cities and Society, Elsevier, volume 55, page-102018, year-2020.
[58] , Li, Wenwen and Batty, Michael and Goodchild, Michael F "Real-time GIS for smart cities", Taylor & Francis, year-2020
[59] , Arjun, R and Suprabha, KR, "Innovation and Challenges of Blockchain in Banking: A Scientometric View", International Journal of Interactive Multimedia & Artificial Intelligence, volume 6, number 3, 2020.

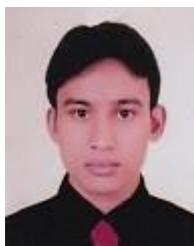

MD. JAHIDUL ISLAM received the B.Sc. and M.Sc. degrees in Computer Science and Engineering from Jagannath University (Jnu), Dhaka, in 2015 and 2017 respectively. Currently, he is working as a Lecturer and Program Coordina- tor (Day) at Computer Science and Engineering (CSE), Green University of Bangladesh (GUB), Dhaka, Bangladesh since May 2017 to present. He is a member of Computing and Communication and Human-Computer Interaction (HCI) research groups, CSE, GUB. His research interests include Internet of Things (IoT), Blockchain, Network Function Virtualization (NFV), Software Defined Networking (SDN), 5G, Digital Forensic Investigation (DFI), HCI, and Wireless Mesh Networking (WMN).

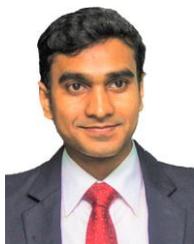

ZIAUR RAHMAN is currently a PhD Candidate at RMIT University, Melbourne, and an associate professor (currently in study leave) of the Department of ICT, MBSTU, Bangladesh. He was graduated from Shenyang University of Chemical Technology, China, in 2012 and completed Masters from IUT, OIC, in 2015. His articles received the best paper award and nomination in different international conferences and published in reputed journals. Also, he is an IEEE and ACM Graduate Member, Australian Computer Society (ACS) Associate Member as well. His research includes Blockchain, Industrial Internet of Things (IIoT), Cybersecurity, and Software Engineering.

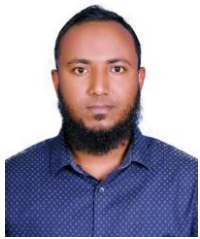

MD. MAHFUZ REZA received his B.Sc.(Engg.) in Computer Science and Engineering (CSE) from Mawlana Bhashani Science and Technology University (MBSTU), Tangail, Bangladesh in 2011. Also, he received M.Sc.(Engg.) in CSE from MBSTU in 2015. Currently, he is an Associate Professor of CSE, MBSTU. His current research interests in the areas of data sciences, machine learning, sensor networks, IoT, cryptography and network security.

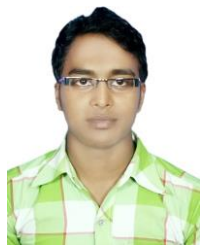

ANICHUR RAHMAN received the B.Sc. and M.Sc degree in Computer Science and Engineering from Mawlana Bhashani Science and Technology University (MBSTU), Tangail, Bangladesh in 2017 and 2020 respectively. Currently, he is working as a Lecturer at Computer Science and Engineering (CSE), National Institute of Textile Engineering and Research (NITER), Savar, Dhaka, Bangladesh since January 2020 to present. His research interests include Internet of Things (IoT), Blockchain (BC), Software Defined Networking (SDN), Image Processing, Machine Learning, Vehicular Ad-Hoc Networking (VANET), 5G, Industry 4.0 and Data Science.

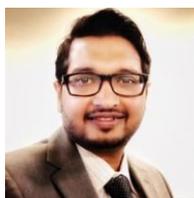

ADNAN ANWAR is a Lecturer and Deputy Director of postgraduate cybersecurity studies at the School of Information Technology. Previously he has worked as a Data Scientist at Flow Power. He has over 8 years of research, and teaching experience in universities and research labs including NICTA, La Trobe University, and University of New South Wales. He received his PhD and Master by Research degree from UNSW. He is broadly interested in the security research for critical infrastructures including Smart Energy Grid, SCADA system, and application of machine learning and optimization techniques to solve cyber security issues for industrial systems. He has been the recipient of several awards including UPA scholarship, UNSW TFR scholarship, best paper award and several travel grants including ACM and Postgraduate Research Student Support (PRSS) travel grants. He has authored over 40 articles including high-impact journals (mostly in Q1), conference articles and book chapters in prestigious venues. He is an active member of IEEE for over 9 years and serving different committees.

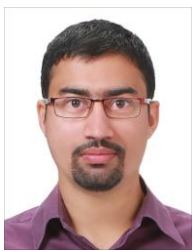
M. A. PARVEZ MAHMUD received his B.Sc. degree in Electrical and Electronic Engineering and Master of Engineering degree in Mechatronics Engineering. After the successful completion of his Ph.D. degree with multiple awards, he worked as a Postdoctoral Research Associate and Academic in the School of Engineering at Macquarie University, Sydney. He is currently an Alfred Deakin Postdoctoral Research Fellow at Deakin University. He worked at World University of Bangladesh (WUB) as a 'Lecturer' for more than 2 years and at the Korea Institute of Machinery and Materials (KIMM) as a 'Researcher' for about 3 years. His research is focused on Energy Sustainability, Secure Energy Trading, Microgrid Control and Economic Optimization, Machine Learning, Data Science, and Micro/nanoscaled Technologies for Sensing and Energy Harvesting. He accumulated experience and expertise in machine learning, life cycle assessment, sustainability and economic analysis, materials engineering, microfabrication, and nanostructured energy materials to facilitate technological translation from the lab to real-world applications for the better society. He has produced over 50 publications, including 1 authored book, 3 Book Chapters, 29 Journal Papers, and 21 fully refereed Conference Papers. He received several awards including "Macquarie University Highly Commended Excellence in Higher Degree Research Award 2019". He was involved in teaching engineering subjects in the Electrical, Biomedical and Bechatronics Engineering courses at the School of Engineering, Macquarie University for more than 2 years. Currently, he is involved in the supervision of 6 PhD students at Deakin University. He is a key member of Deakin University's Advanced Integrated Microsystems (AIM) research group. Apart from this, he is actively involved with different professional organizations, including Engineers Australia and IEEE.

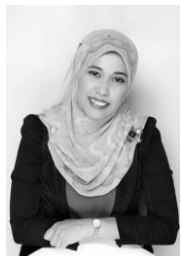
RAFIDAH MD. NOOR received the bachelor's degree in IT from University Utara Malaysia, in 1998, the M.Sc. degree in computer science from Universiti Teknologi Malaysia, in 2000, and the Ph.D. degree in computing from Lancaster University, U.K., in 2010. She is currently an Associate Professor with the Department of Computer Systems and Technology, Faculty of Computer Science and Information Technology, University of Malaya, and the Director of the Centre of Mobile Cloud Computing Research, which focuses on high-impact research. She has performed nearly RM 665 606.00 for High-Impact Research, Ministry of Education Grant, and other research grants from the University of Malaya and public sectors. She has supervised more than 30 postgraduate students within five years. She has published more than 50 articles in Science Citation Index and Expanded Non-Science Citation Index. The proceeding articles were published in international/national conferences and in a few book chapters. Her research is related to transportation systems in computer science, including vehicular networks, wireless networks, network mobility, quality of service, and the Internet of Things.

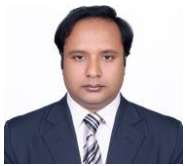
MOSTOFA KAMAL NASIR Professor of Computer Science and Engineering of Mawlana Bhashani Science and Technology University, Tangail, Bangladesh. He has completed his PhD from University of Malaya, Kuala Lumpur, Malaysia in the field of Mobile Adhoc Technology in 2016. Before that he has completed his BSc and MSc in Computer Science and Engineering from Jahangirnagar University, Bangladesh.